\documentclass[twocolumn,showpacs,preprintnumbers,amsmath,amssymb,floatfix,prb]{revtex4}

\usepackage{graphicx}
\usepackage{dcolumn}
\usepackage{bm}

\begin{document}


\title{Surfactant-like Effect and Dissolution of Ultrathin Fe Films 
on Ag(001)
}
\author{S. Terreni$^{a}$}
\author{A. Cossaro$^{b}$}
\author{G. Gonella$^{a}$}
\author{L. Mattera$^{a}$}
\author {L. Du\`o$^{c}$}
\author{F. Ciccacci$^{c}$}
\author {D. Cvetko$^{b}$}
\altaffiliation[Also at ]{Physics Department, University of  
Ljubljana, Slovenia.}
\author {L. Floreano$^{b}$}   
\author {A. Morgante$^{b}$}
\altaffiliation[Also at ]{Physics Department, University of Trieste, 
Italy.}
\author {A. Verdini$^{b}$}
\author{M. Canepa$^{a}$}
\email{Canepa@fisica.unige.it}
\affiliation{$^{a}$Unit\`{a} INFM and CNR-IMEM, Dipartimento di 
Fisica, Universit\`{a} di Genova, via Dodecaneso 33, I-16146 Genova, 
Italy}
\affiliation{$^{b}$Laboratorio TASC-INFM, Trieste, Italy}
\affiliation{$^{c}$INFM and Dipartimento di Fisica, Politecnico di 
Milano, 
Italy }

\date{\today}

\begin{abstract}
The effects of annealing on the structure of ultra thin  Fe films (4 
- 10 ML) 
deposited at 150 K on Ag(001) were studied by synchrotron radiation
photoelectron diffraction (PED) and x-ray diffraction (XRD).
The occurrence of a surfactant-like stage, in which a single layer of 
Ag  covers the Fe film is 
demonstrated for films of 4-6 ML heated at 500-550 K. Evidence of a 
stage 
characterized by the formation of two Ag capping layers is also 
reported. 
As the annealing temperature was increased beyond 700 K the surface 
layers closely resembled 
the structure of bare Ag(001) with the residual presence of 
subsurface Fe aggregates.  
The data illustrate a film dissolution path which is in agreement 
with recent theoretical models [J. Roussel 
et al. Phys. Rev. B {\bf 60}, 13890 (1999)]. 
\end{abstract}


\maketitle

The phase immiscibility and the excellent matching between Ag(001) 
and Fe(001) unit cells
 (mismatch  0.8~$\%$) make Fe/Ag growth attractive in the field of 
low dimensionality magnetic  systems, 
such as ultrathin films,\cite{ultrathin}  
multilayers,\cite{multilayers}  
and small aggregates \cite{nogueira}. 
At the nanometric scale, atomic exchange 
processes were found to affect the chemical 
sharpness of interfaces in films and multilayers.\cite{egelhoffmrs} 
Intermixing could be drastically limited at deposition temperatures 
as 
low as 140-150~K \cite{egelhoffmrs,lille,langelaar} at the expense of 
a poor morphological quality of the 
film.\cite{prbrocking,burgler}
The film structural evolution induced by post-growth annealing 
presents many 
interesting aspects involving activated atomic exchange processes and 
affecting  
magnetic properties.\cite{magnetism} Previous experiments, of He 
and low energy ion scattering on films deposited at 150 K, 
indicated the formation of a segregated Ag layer upon
moderate annealing (550~K). Higher temperatures led to the embedding 
of 
Fe into the Ag matrix.\cite{ricciardi} 
In those experiments, information on sub-surface layers was attained 
by ion erosion depth 
profiling, a destructive technique mainly sensitive to the 
topmost layer.   Many questions remained open 
about the film structure and morphology evolution during dissolution.
Here we address this issue by presenting  photoelectron and x-ray 
diffraction 
experiments, 
performed at the ALOISA beam line (ELETTRA, Trieste).
PED provides chemically selected data on film structure with an 
information depth of 
several layers. It allows to characterize local order in films which 
are disordered on a long range scale and possesses 
specific sensitivity to segregation 
processes. \cite{pedgenerale}
Systematic PED measurements have been accompanied with a few XRD 
rod scans yielding a better sensitivity to the buried interface and 
the film long range order. 
The results of this paper allow a comparison with recent 
models enlightening the dissolution 
paths of an ultra thin metal film into a different metal, 
when both subsurface 
migration of the deposit  and phase separation 
between substrate and deposit are 
favoured.\cite{treglia1ML,treglia10ML}

Details on the ALOISA system can be found elsewhere.\cite{aloisa} 
The Ag substrate was prepared according to well established 
procedures.\cite{prbrocking}  
Reflection High Energy Electron Diffraction was used to monitor 
the surface order.
XPS surveys were employed to 
check the chemical composition of the surface. Grazing incidence  
XPS was also used to monitor 
in real time Fe {\it 2p} and Ag {\it 3p} signals during annealing 
ramps. 
The sample temperature was controlled by thermocouples and by an 
optical pyrometer. 
Guided by results of refs. \cite{egelhoffmrs,ricciardi}, heating was 
stopped 
at progressively higher temperatures, then cooling the system and 
looking at film structure. 
Iron was evaporated by electron bombardment, controlling  
the deposition flux ($\sim$~1.5~\AA/min) by a quartz microbalance.
Measurements were focused on films in the 4-10 ML thickness range 
deposited at  150~K.
Grazing incidence XRD have been applied on  a few films by
measuring radial scans across the $(\overline{2} 0 0)$ and $(2 
\overline{2} 0)$ peaks in the in--plane 
Ag(001) reciprocal lattice. These measurements yielded the lateral 
lattice 
spacing with a precision better than 0.01~\AA.
The vertical structure of the Fe film has been also probed by 
out-of-plane 
XRD (rod scan), taken for the $(\overline{2} 0 L)$ rod at a photon 
energy of 6000 eV. 
PED polar scans were measured at grazing incidence 
(h$\nu$~=~900-1300~eV) 
in Transverse Magnetic polarization,  by rotating an electron energy 
analyzer in 
the scattering  plane. 
The notation Ag (Fe)~$nl(\theta )$,  
will indicate  PED scans as a function of the polar angle $\theta$ 
from the surface normal, obtained by photoemission 
from $nl$ states of Ag (Fe). The photoemission intensity was 
collected at the peak energy and at suitably chosen 
energies along the peak tails in order to allow for a subtraction 
of the background due to secondary electrons. FF will denote forward 
focusing peaks along off normal 
nearest neighbour atomic chains. For the sake of synthesis we will 
show 
only data taken with the $\langle 001 \rangle$ Ag surface direction 
in the scattering plane.  
 
\begin{figure}[tbp]
\includegraphics[width=8.5cm]{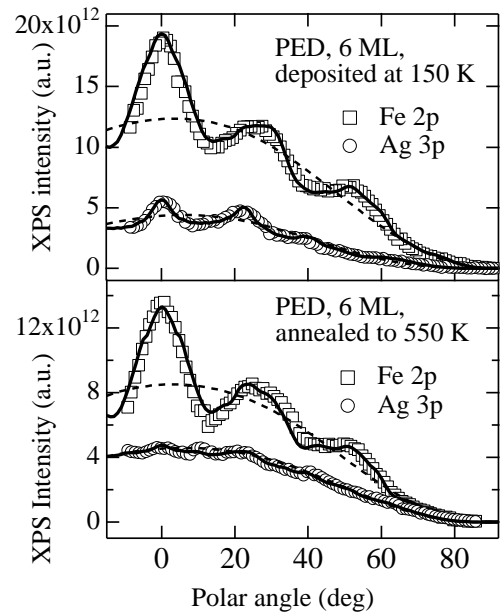}
\caption{\label{fig1}Fe {\it 2p} $(\theta )$ and  Ag {\it 3p} 
$(\theta )$ 
PED patterns obtained on a 6$\pm 0.5$  ML Fe film (upper panel) 
deposited 
at 150 K and (lower panel) after 
annealing at 550 K.  Measurements were taken at 150 K along the 
$\langle100\rangle$ azimuth of the Ag substrate.  
In both panels continuous and dashed lines represent the calculated 
I$_{calc}$($\theta$) and ISO$_{calc}$ patterns, 
respectively.}
\end{figure}

The computational approach to PED was thoroughly described 
elsewhere.\cite{isobruno} 
In brief, the polar scans  I$_{exp}$($\theta$) have been compared 
with calculated 
I$_{calc}$($\theta$)~=~ISO$_{calc}$($\theta$)~$(1~+~\chi_ 
{calc}$)($\theta$).
The anisotropy term $\chi_ {calc}$($\theta$), carrying information on 
the arrangement of atoms 
around 
the photo-emitter, was taken proportional to the output generated by 
the Multiple Scattering Calculation of Diffraction 
(MSCD)  
package.\cite{mscd} 
Several structural models have been considered both for the 
as-deposited and the annealed films.
We varied the number of Fe planes and considered the presence of one 
or more Ag layers over the Fe film; 
intermixed interfaces were also examined.
In all calculations, the in-plane lattice constants 
of the film $a_{Fe}$ was set to the substrate value 
$a_{Ag}$~=~2.88~\AA, 
inferred from in-plane grazing incidence XRD, 
which indicated  a pseudomorphic in-plane structure, 
independently on the investigated thickness and annealing 
temperature. 
In fact, the formation of a non pseudomorphic Fe film would have 
yielded side-peaks 
or shoulders close to the substrate in-plane XRD 
peaks,\cite{prbbruno} which were 
never observed.
The isotropic term  ISO$_{calc}$($\theta$) has been calculated, 
taking into account emission matrix elements, 
electron escape path, surface roughness and  
instrumental factors.\cite{isobruno}
Regarding as-deposited films, 
systematic PED (and Auger electron diffraction) measurements, taken 
along several azimuthal orientations, indicated growth of  Fe films 
with body centered cubic 
structure and unit cell rotated by 45$^\circ$ with respect to the 
substrate 
cell (in brief $^{R45}$~bcc structure). 
The body of data resulted in close agreement with 
literature.\cite{egelhoffmrs,litonner}
For films thicker than 5 ML the ratio between  the vertical and 
in-plane lattice constants  
$c_{Fe}/a_{Fe}$ matched unity, as expected for an ideal bcc 
structure. On the thinnest films the FF peak was 
slightly shifted from the bcc position,  giving a value 
$c_{Fe}/a_{Fe}$~=~1.06~$\pm$~0.01 for a 4 ML film. 
This expansion of $c_{Fe}$,  
was early attributed to the influence of intermixing at the 
interface\cite{egelhoffmrs,prbrocking}.
Representative measurements on a 6 ML film are reported in the upper 
panel of Fig.~\ref{fig1}. 

In the figure, the calculations are related to a model with 6 layers 
of 
Fe ($^{R45}$~bcc structure)
laying on the substrate mimicked by 3 layers of 
Ag (in brief 6Fe/Sub). Concerning the Fe pattern, the simulation 
was insensitive to the value of the 
interlayer distance at the interface ($d_{Ag}^{Fe}$).
The best simulation of the Ag {\it 3p } $(\theta)$ pattern, showing a 
strong attenuation 
of intensity and significant shape 
variations with respect to the bare Ag(001), was obtained 
with $d_{Ag}^{Fe}$~=~1.70~$\pm$~0.03~\AA. 
This value, in between Ag ($d_{Ag}$~$\sim$~2.04~\AA) and Fe 
($d_{Fe}$~$\sim$~1.43~\AA)  bulk 
interlayer spacings, enabled reproduction 
of the main features of the pattern, including  the suppression of 
the FF peak 
typical of fcc structure,  at 45$^\circ$.
 
Fe {\it 2p} $(\theta )$ and  Ag {\it 3p} $(\theta )$ patterns 
obtained on the 
same 6 $\pm 0.5$ ML Fe film after annealing  at 550 K are shown in 
the lower panel 
of Fig.~\ref{fig1}. 
The shape of the Fe PED pattern was  similar to the as-deposited one. 
The Ag {\it 3p }$(\theta)$ data,  showing a quasi isotropic 
behaviour  suggested the 
simple physical model of a Ag overlayer.
Continuous lines in the figure were calculated  according to a 
1Ag/6Fe/Sub scheme.
The structure of the Fe film was the same used for the as-deposited 
film. 
The two Fe/Ag interfaces were assumed sharp.  
The Fe/Ag distance at 
the top layer resulted $d_{Fe}^{Ag}$~=~1.70~$\pm$~0.03~\AA.  
The simulations reproduce rather accurately both Fe and Ag PED 
patterns supporting quantitatively the 
assumed structural model.
  
\begin{figure}[tbp]
\includegraphics[width=8.4cm]{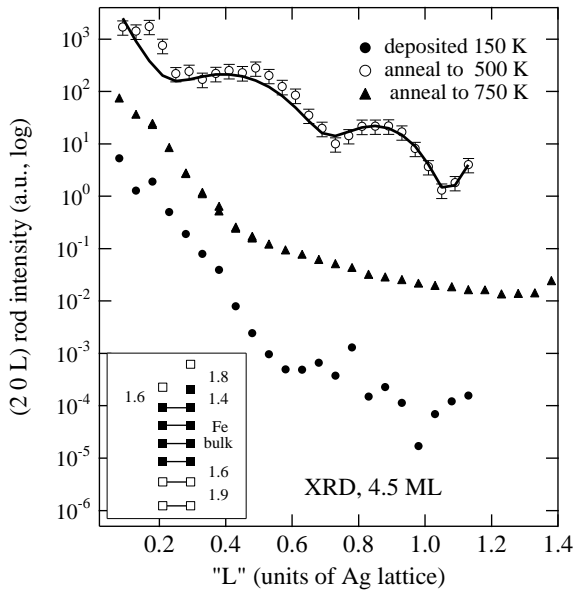}
\caption{\label{fig2}
XRD scan of the $(\overline{2} 0 L)$  rod of Ag 
for a   5$\pm 0.5$~ ML Fe film. Measurements 
taken after the deposition  and after annealing 
at different temperatures  are shown. 
Modulations in the upper pattern  arise from the 
formation of sharp interfaces at both top and bottom of the film. 
Fitting to the data (thick full line) yielded the structural model 
sketched 
in the inset, where filled and open squares represent the Fe and Ag 
layers, respectively. The topmost Ag layer is half filled, and the 
layer beneath 
is formed by 1/2~ML of Fe and 1/2~ML of Ag. Best fit layer spacings 
are also reported (in \AA).}
\end{figure}

Out-of-plane XRD measurements taken on a 5 $\pm 0.5$ ~ML Fe film are 
consistent 
with the model drawn from PED. The XRD scans along the $(\overline{2} 
0 L)$ rod are shown 
in Fig.~\ref{fig2}, as taken for the just deposited (150~K) film and  
after annealing at 500~K and 750~K. The absence of regular features 
in the pristine 
film was attributed to the lack of long range order and to the large 
surface 
roughness. After annealing to 500~K, well defined modulations 
appeared, which witnessed 
the formation of large Fe domains and sharp interfaces. 
Fitting to the rod scan of the annealed film yielded a model with 
4.5  
Fe layers covered by one surfactant 
Ag layer. 
Due to the reduced region of reciprocal space presently explored, 
we have reduced the fitting parameters by considering a simple model 
where the three inner Fe layers were
fixed to the bulk spacing. In fact, the rod modulations are mainly 
sensitive to 
the film interfaces, where we focussed our fitting optimization. 
We found an height of 1.6 and 1.8~\AA~ for the Ag surfactant layer 
above the 4th and 5th Fe layer, respectively.
This discrepancy is simply due to the model simplification, which 
does 
not allow the 4th Fe layer to relax (see inset of Fig.~\ref{fig2}).
The width of the buried Fe-Ag interface was found to be 1.6~\AA.
The height of the first Ag layer below the Fe film was contracted to 
1.9~\AA. 
Same quality fits were obtained by admitting intermixing 
at the first Ag layer below the Fe film (10-15\%). In this case the 
buried interface width was also slightly affected (uncertainty of 
0.1~\AA).

\begin{figure}[tbp]
\includegraphics[width=8.5cm]{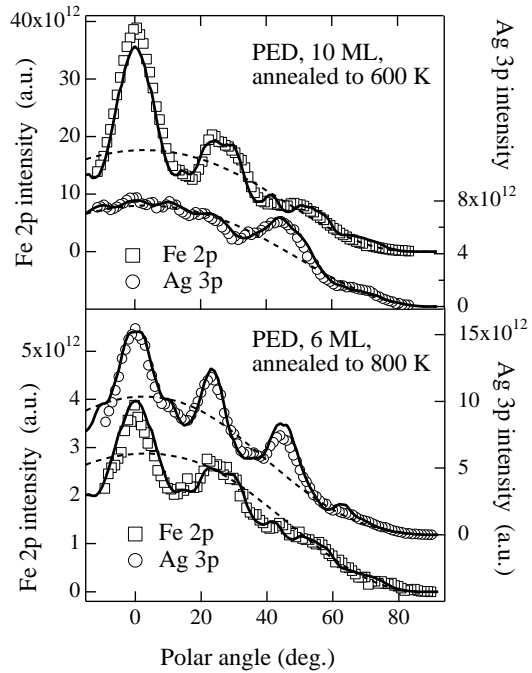}
\caption{\label{fig3}
Upper panel: Fe${\it 2p} (\theta )$ (Squares) and  Ag${\it 2p} 
(\theta )$ 
(Circles) PED patterns on a 10~$\pm$~0.5~ML Fe 
film deposited at 150~K and annealed to 600~K. Measurements were 
taken at 
170~K along the 
$\langle100\rangle$ azimuth of the Ag substrate. Fe and Ag intensity 
are reported in the left and right axis, respectively.
Continuous and dashed lines represent the 
calculated I$_{calc}$($\theta$) and ISO$_{calc}$ 
patterns, respectively, as described in the text.
Lower panel: as in the upper panel, but for
a 6 $\pm 0.5$ ML Fe film after annealing at 800~K.}
\end{figure}

In the upper panel of Fig.~\ref{fig3}, we show PED data obtained on a 
10$\pm 0.5$ ML Fe film measured after 
 annealing at 600 K. Fe {\it 2p} $(\theta)$ continued to show 
bcc-like 
structure, but with an increased anisotropy  along the normal 
direction.
The Ag {\it 3p }$(\theta)$ data show  faint 
features at 0$^\circ$, $\sim$~10$^\circ$,
$\sim$~20$^\circ$ and a huge
peak at 44$^\circ$ , indicating the formation 
of two Ag surfactant layers.
Addition of a second Ag surface layer together with   
optimization of the $d_{Fe}^{Ag}$  distance at the interface and of 
the $d_{Ag}^{Ag}$ distance between 
the first and second Ag layer, yielded a satisfactory simulation of 
the Ag PED pattern.
The I$_{calc}$ curves, superimposed to the data of Fig.~\ref{fig3}, 
were calculated considering  
a 2Ag/7Fe model with  $d_{Fe}^{Ag}$~=~1.73 $\pm 0.03$~\AA~  and 
$d_{Ag}^{Ag}$~=~1.90 $\pm 0.03$~\AA.  
$d_{Fe}^{Fe}$ was set at 1.43~\AA.
The simulation of the Fe {\it 2p} data was less satisfactory 
suggesting that the model 
oversimplifies the physical situation, where the second Ag layer 
could 
be only partially filled.

Upon further increasing the annealing temperature, the Ag {\it 3p} 
PED patterns 
gradually approached the one obtained 
on bare Ag(001).  
Beyond $\sim$~650~K  we could 
not find an unique geometrical input able to describe both Fe {\it 
2p} and Ag {\it 3p} PED patterns. 
Representative data obtained on the 6 $\pm 0.5$ ML film after 
annealing at 800~K
are shown in the lower panel of Fig.~\ref{fig3}.
The  Ag pattern was  reproduced considering the same structural model 
used 
to fit the PED taken on the clean substrate (not shown).  
On the other hand models with Fe layers buried under many Ag layers, 
could not fit Fe {\it 2p} patterns at all.
The overall intensity  of the Fe {\it 2p} pattern presented a  
relevant attenuation with respect to the as-deposited film. However,  
a significant anisotropy is still appreciable. 
These findings suggested the presence of 
subsurface Fe aggregates of small lateral size.
Similar  effects were also observed from XRD, where the rod scan 
taken after annealing to 750~K (filled triangles in Fig.~\ref{fig2})
yielded a smooth decay due to the 
disappearance of the subsurface Fe film. This rod scan is indicative 
of either Fe dissolution in the Ag matrix either fragmentation of the 
Fe film into small clusters of irregular size and spatial 
distribution.
In order to get more information on such aggregates from PED data, we 
explored 
simple $N$Ag/$M$Fe  
structural models in which
$M$ layers of Fe (for simplicity with bcc structure) were buried 
below $N$ layers of Ag  
(with interlayer distance 
$d_{Ag}^{Ag}$). A sharp Fe/Ag interface  was considered with 
interface spacing 
$d_{Fe}^{Ag}$.   
 In the case $M$ = 6, i.e. assuming for aggregates the same nominal 
thickness 
 of the as-deposited film,  
a qualitative agreement (not shown) could be obtained with 
$N$ = 1 or 2. The agreement became definitely worse with 
larger values of $N$.
A fair reproduction of the pattern (including the twinned  PED 
feature 
at about 45$^\circ$ and 55$^\circ$) could be finally 
obtained with a 2Ag/8Fe model, 
$d_{Ag}^{Ag}$ = 1.87 $\pm$ 0.03 \AA, $d_{Fe}^{Ag}$ = 1.67 $\pm$ 0.03 
\AA~  
(full lines in Fig.~\ref{fig3}). 
Thus, although the shape and size of the Fe subsurface aggregates 
could not be determined, 
the analysis of PED 
pattern confirms, on a quantitative ground, 
the residual presence of subsurface Fe clusters. 
 
Our data can be rationalized in the light of recent theories  on 
dissolution of 
ultra thin metal film into a metal substrate  which enlightened 
different routes depending on 
the physico-chemical coupling of the deposit (A) and substrate (B) 
\cite{treglia1ML,treglia10ML}. 
When A and B present a tendency to bulk ordering, the dissolution can 
be blocked by the 
formation of AB surface alloys. 
In case of tendency to A/B phase separation, dissolution paths will 
critically depend 
on the A and B surface energies. 
The case of interest here, in which B atoms show a stronger tendency 
to surface segregation 
than A atoms as for Fe/Ag or Ni/Ag systems,\cite{aufray} was 
illustrated 
in Ref.\cite{treglia10ML} on the example of a  
10 ML deposit
(schematically indicated as A/A/$\ldots$/A/B). According to 
calculations, 
if the temperature is raised beyond a threshold T$_C$, 
dissolution proceeds through the 
so called surfactant-layer-by-layer (SLBLD) mode. 
In this regime, on a short range time scale, the system passes 
through a surfactant like stage, with formation of successive 
B/A/A/$\ldots$/A/B,  
B/B/A/A/$\ldots$/A/B and B/B/B/A/A/$\ldots$/A/B profiles. 
The excellent agreement between data of figs.~\ref{fig1} and 
\ref{fig2} with 
a 1Ag/Fe/Fe/$\ldots$/Fe/Sub model 
nicely demonstrates the occurrence of the 
B/A/A/../A/B stage. Note that the MSCD simulations provide a reliable 
indication that the local 
structure of the Fe film remains substantially intact, in agreement 
with the model  
which predicts a negligible loss of deposit matter into the bulk at 
this stage.\cite{treglia10ML}
In addition, XRD rod scans indicate that the formation of the first 
Ag 
surfactant layer does not involve a fragmentation of the Fe film. 
Rather the film is ordered to a uniform thickness in the early 
annealing stage and this order is preserved during the segregation 
of the first surfactant layer.
Data of Fig.~\ref{fig3} (upper panel), representing a physical 
situation 
mimicking the  B/B/A/$\ldots$/A/B stage, 
add further confidence about the agreement of our results with the 
model.
On the basis of these results, we feel encouraged to extend the 
conclusions about the 
surfactant stage to the larger spatial scale probed by He diffraction 
in previous experiments performed under similar experimental 
conditions.\cite{ricciardi}       
Data of Fig.~\ref{fig3} (lower panel) illustrate a clear trend 
towards film dissolution,
which appears in qualitative agreement with the final 
stage predicted by theory, i.e. total disappearance of deposit 
species 
down to the tenth substrate layer, for T $>$ T$_C$.
Residual Fe clusters in the PED data for the highest annealing 
temperature
is the only deviation of our data  from the SLBLD model, possibly due 
to specific details of the present system surface energies or to 
experimental limitations.  
For example, Oxygen impurities tend to 
locally lower the surface energy of the film, 
therefore acting as surfactant species\cite{bonanno} 
competing with Ag.

\end{document}